\documentclass[12pt,a4paper]{article}
\usepackage{a4wide}
\usepackage{latexsym}
\usepackage{epsf}
\usepackage{amssymb}

\usepackage{amsmath, cite}
\usepackage{slashed,epsfig}

\makeatletter
\@addtoreset{equation}{section}
\makeatother

\newsavebox{\uuunit}
\sbox{\uuunit}
    {\setlength{\unitlength}{0.825em}
     \begin{picture}(0.6,0.7)
        \thinlines
        \put(0,0){\line(1,0){0.5}}
        \put(0.15,0){\line(0,1){0.7}}
        \put(0.35,0){\line(0,1){0.8}}
       \multiput(0.3,0.8)(-0.04,-0.02){12}{\rule{0.5pt}{0.5pt}}
     \end {picture}}

\begin{document}

\begin{flushright}
\small
CERN-TH/2003-133\\
IFT-UAM/CSIC-03-21\\
UG-03-03\\
{\bf hep-th/0306179}\\
\date \\
\normalsize
\end{flushright}

\begin{center}


\vspace{.7cm}

{\LARGE {\bf The Bianchi Classification of}}
\bigskip
\bigskip

{\LARGE {\bf Maximal $D=8$ Gauged Supergravities}}

\vspace{1.2cm}

{\large Eric Bergshoeff}${}^{\diamondsuit}$
,
{\large Ulf Gran}${}^{\diamondsuit}$
, 
{\large Rom\'an Linares}${}^{\diamondsuit}$
, \\
\smallskip
\smallskip

{\large Mikkel Nielsen}${}^{\diamondsuit}$
,
{\large Tom\'as Ort\'{\i}n}${}^{\spadesuit,\clubsuit}$
{\large and}
{\large Diederik Roest}${}^{\diamondsuit}$
\vskip 1truecm

\small
${}^{\diamondsuit}$\ {\it Centre for Theoretical Physics, University of 
Groningen,\\
   Nijenborgh 4, 9747 AG Groningen, The Netherlands\\
E-mail: {\tt (e.a.bergshoeff, u.gran, r.linares, m.nielsen, 
d.roest)@phys.rug.nl}}
\vskip 0.3cm
${}^{\spadesuit}$\ {\it Instituto de F\'{\i}sica Te\'orica, C-XVI,
Universidad Aut\'onoma de Madrid \\
E-28049-Madrid, Spain}
\vskip 0.3cm
${}^{\clubsuit}$\ {\it Theory Division, C.E.R.N., CH-1211, Geneva 23, 
Switzerland\\
E-mail: {\tt Tomas.Ortin@cern.ch}}

\vspace{.7cm}


{\bf Abstract}

\end{center}

\begin{quotation}

\small

We perform the generalised dimensional reduction of $D=11$ supergravity over 
three-dimensional group manifolds as classified by Bianchi. Thus, we construct 
eleven different maximal $D=8$ gauged supergravities, two of which have an 
additional parameter. One class of group manifolds (class B) leads to 
supergravities that are defined by a set of equations of motion that cannot be 
integrated to an action.

All 1/2 BPS domain wall solutions are given. We also find a 
non-supersymmetric domain wall solution where the single transverse 
direction is time. This solution describes an expanding universe and upon 
reduction gives the Einstein-de Sitter 
universe in $D=4$. The uplifting of the different solutions to 
M-theory and the isometries of the corresponding group manifold are discussed.

\end{quotation}

\newpage

\pagestyle{plain}

\section{Introduction}

The first example of a maximal $D=8$ gauged supergravity is the $SO(3)$
gauged supergravity constructed in 1985 by Salam and Sezgin \cite{Salam:1985ft}.
In recent years this maximal $D=8$ gauged supergravity has regained 
interest for several reasons. First of all the $D=8$ theory
was used in the construction of the dyonic membrane \cite{Izquierdo:1996ms}.
It also occurs in the DW/QFT correspondence when one considers the
near-horizon limit of the D6-brane \cite{Boonstra:1998mp}.
Soon after, a number of papers appeared where maximal $D=8$ gauged 
supergravity 
played an important role in the construction of special holonomy manifolds
by considering wrapped branes, see 
e.g.~\cite{Edelstein:2001pu,Hernandez:2001bh,Gomis:2001vg,Hernandez:2002fb}.
More recently, the same theory also turned up in a discussion
of gravitational topological quantum field theories \cite{Baulieu:2003gq}
and accelerating universes \cite{Burgess:2003mk}.

In view of all these applications, it is of interest to ask oneself how unique 
the $SO(3)$ gauged supergravity is. In a recent paper 
\cite{AlonsoAlberca:2003jq}, we performed a generalised dimensional reduction 
of $D=11$ supergravity over three-dimensional group manifolds 
\cite{Dewitt,Scherk:1979zr}\footnote{The construction of gauged maximal 
supergravities (with a Lagrangian) in diverse dimensions is discussed from a 
purely group-theoretical point of view in \cite{deWit:2002vt}.}. By using one 
class of group manifolds, called class A in \cite{refc}, we constructed five 
other maximal $D=8$ supergravities with gauge groups $SO(2,1), ISO(2), 
ISO(1,1)$, Heis$_3$ and $U(1)^3$. Here Heis$_3$ denotes the three-dimensional 
Heisenberg group (with generators corresponding to position, momentum and 
identity). The theory with gauge group $U(1)^3$ is obtained from a reduction 
over a torus $T^3$ and is referred to as the ungauged theory, since there are 
no fields that carry any of the $U(1)$ charges. All groups mentioned above are 
related to $SO(3)$ by group contraction and/or analytic continuation. We will 
refer to them as class A supergravities.

In the same paper we showed that by using another class of group manifolds, 
called class B in \cite{refc}, yet more gauged supergravities can be 
constructed whose gauge groups can be seen as extensions of $ISO(2), 
ISO(1,1)$, Heis$_3$ and $U(1)^3$. We call these class B supergravities. There 
is an extensive literature on the fact that a class B group manifold reduction 
leads to inconsistent field equations when reducing the action, a fact first 
noticed by Hawking \cite{refe} and discussed in \cite{refd} (recent overviews 
are given in \cite{refk,Cvetic:2003jy}). This is related to the fact that the 
field equations following from the reduced action do not coincide with the 
reduction of the field equations themselves \cite{Pons:1998tt}. We find, by 
explicitly performing the group manifold procedure, that the reduction can be 
performed on-shell, i.e.~at the level of the equations of motion or the 
supersymmetry transformations. Particularly in string theory this seems to be 
a relevant approach since the world-sheet theory yields space-time field 
equations rather than an action principle.

Making use of the fact that the class B group manifolds have
two commuting isometries, we provide an alternative way of viewing this
issue by relating the class B group manifold reduction to a 
Scherk-Schwarz reduction \cite{Scherk:1979ta} from nine dimensions. 
In this procedure one uses an internal scale symmetry that leaves the $D=9$ equations of motion 
invariant but scales the Lagrangian. We indicate the M-theory origin of this 
scale symmetry. The fact that such a scale symmetry of the equations of motion 
can be used for a Scherk-Schwarz reduction and leads to equations of motion 
that cannot be integrated to a Lagrangian was first observed in 
\cite{Lavrinenko:1998qa}.

Supergravities without
an action naturally occur in the free differentiable algebra approach
to supergravity, see e.g.~\cite{Andrianopoli:2000fi}. Furthermore, they
have occurred recently in a study of matter-coupled
supergravities in $D=5$ dimensions \cite{Bergshoeff:2002qk}. We note that,
although the $D=8$ class B supergravities have no Lagrangian, there
is a hidden Lagrangian in the sense that these theories can be obtained
by dimensional reduction of a theory in $D=11$ dimensions
with a Lagrangian. 

Note that among the different group manifolds there is a number of non-compact 
manifolds, in particular all class B group manifolds. Thus, many of the 
reductions we perform are not compactifications in the usual sense (on a small 
internal manifold) but rather consistent truncations of the full 
higher-dimensional theory to a lower-dimensional subsector. The consistency of 
the truncation guarantees that $D=8$ solutions uplift to $D=11$ solutions. We 
do not consider global issues here and focus on local properties.

In \cite{AlonsoAlberca:2003jq}, we showed that for the five maximal
$D=8$ class A gauged supergravities, the most general
domain wall solution is given by a so-called $n$-tuple domain wall solution 
with $n \leq 3$, 
which can be viewed as the superposition of $n$ domain walls. The
embedding of this $n$-tuple domain wall solution into M-theory for the 
$SO(3)$ case naturally includes the near-horizon limit of the Kaluza-Klein 
monopole,
as conjectured by the DW/QFT correspondence.
In \cite{Boonstra:1998mp}, the $SO(3)$
domain wall with $n=1$ was found to correspond to the
one-centre monopole, while we found in \cite{AlonsoAlberca:2003jq} that 
the $SO(3)$ domain wall with $n=2$
uplifted to the near-horizon limit of the two-centre 
monopole solution.
Both these solutions are non-singular upon uplifting whereas the 
$n=3$ domain wall uplifts to a singular space-time \cite{Belinskii:1978}.

In this paper we continue the work of \cite{AlonsoAlberca:2003jq}. 
In particular, we construct the supersymmetry rules of maximal 
$D=8$ gauged supergravities for both class A and B by dimensionally 
reducing the $D=11$ supergravity over a 
three-dimensional group manifold \cite{Dewitt,Scherk:1979zr}.
In this procedure the three-dimensional Lie algebra defining
the group manifold becomes the algebra of the gauge group after reduction, and 
it is due to this approach that the classification of the 
eight-dimensional gauged supergravities
coincides with the Bianchi classification of three-dimensional 
Lie algebras \cite{Bianchi}.
The reduction gives rise to eleven different maximal $D=8$ gauged 
supergravities, two of which have an additional parameter.
We show how all of these theories, except those whose gauge group is simple 
(i.e.~$SO(3)$ or $SO(2,1)$ for three-dimensional groups), can be obtained by a 
generalised reduction of maximal $D=9$ ungauged supergravity 
using its global symmetry group \cite{Scherk:1979ta}. 

After constructing the different theories, we investigate the 1/2 BPS domain 
wall solutions for both class A and B. We also discuss the isometries of the 
corresponding group manifold and find that the class A $n$-tuple domain wall 
solution of \cite{AlonsoAlberca:2003jq} gives a natural realisation of 
isometry enhancement on a group manifold as discussed in Bianchi's paper 
\cite{Bianchi}. We find that the domain wall solutions of two class A 
supergravities allow for the maximum number of isometries. In addition
we identify the remaining class A solution with six isometries. 

The Bianchi classification suggests that there are two further solutions with maximum number 
of isometries for class B supergravities. We show that these indeed exist and
are given by the same solution. This is a space-like domain wall solution, i.e.~a domain wall 
solution where the single transverse direction is time, and it describes an 
expanding universe with the same qualitative features as the Einstein-de 
Sitter universe. By instead reducing over a seven-dimensional group manifold, 
we find the Einstein-de Sitter universe in $D=4$, which might be an acceptable 
model of our universe \cite{Blanchard:2003du}. The uplifting of the different 
domain wall solutions to M-theory is discussed.

The outline of the paper is as follows. In section 
\ref{Bianchiclassification} we review the Bianchi classification of 
three-dimensional Lie groups and group manifolds. In section \ref{SSreduction} 
we perform the dimensional reduction of $D=11$ supergravity over a 
three-dimensional group manifold, thereby constructing eleven different 
maximal $D=8$ gauged supergravities. In section \ref{D=9} we show that each of 
these theories, except the $SO(3)$ and $SO(2,1)$ cases, can be obtained by a 
generalised reduction of the unique maximal $D=9$ ungauged supergravity. 
Various solutions of these theories are discussed in section \ref{solutions}. 
Our conclusions are given in section \ref{conclusions}. There are two 
appendices. Appendix \ref{ss-rules} gives the supersymmetry rules of the $D=8$ 
theories we construct in this work, while appendix \ref{killingvectors} gives 
the explicit expressions of the Killing vectors corresponding to the isometry 
enhancement of the group manifolds for the Bianchi types relevant for the 
solutions we consider.

\section{Bianchi Classification of 3D Groups and Manifolds}
\label{Bianchiclassification}

In this section we review the Bianchi classification \cite{Bianchi} of 
three-dimensional Lie groups and discuss how these can be realised as 
isometries on three-dimensional Euclidean manifolds\footnote{The 
classification method used nowadays and presented here is not Bianchi's 
original one, but it is due to Sch\" ucking and Behr (see Kundt's paper based 
on the notes taken in a seminar given by Sch\" ucking \cite{refg} and the 
editorial notes \cite{reff}), and the earliest publications in which this 
method is followed are \cite{refb,refc}. The history of the classification of 
3- and 4-dimensional real Lie algebras is also reviewed in \cite{refi}.}.

We assume that the generators of the three-dimensional Lie group
satisfy the commutation relations ($m=1,2,3$)
\begin{align}
  [ T_m , T_n ] = f_{mn}{}^p T_p \,, 
\end{align}
with constant structure coefficients $f_{mn}{}^p$ subject to the Jacobi 
identity $ f_{[mn}{}^q f_{p]q}{}^r = 0$. For three-dimensional Lie groups, the 
structure constants have nine components, which can be conveniently 
parameterized by
\begin{align}
  f_{mn}{}^p = \epsilon_{mnq} Q^{pq} + 2 \delta_{[m}{}^p a_{n]} \,, 
\qquad
  Q^{pq} a_q =0 \,.
\end{align}
Here $Q^{pq}$ is a symmetric matrix with six components, and $a_m$ is a vector 
with three components. The constraint on their product follows from the Jacobi 
identity. Having $a_q=0$ corresponds to an algebra with traceless structure 
constants: $f_{mn}{}^n =0$. Following \cite{refc} we distinguish between class 
A and B algebras which have vanishing and non-vanishing trace, respectively.

Of course Lie algebras are only defined up to changes of basis, $T_m 
\rightarrow R_m{}^n \, T_n$, with $R_m{}^n \in GL(3,\mathbb{R})$. The 
corresponding transformation of the structure constants and its components 
reads
\begin{align}
  f_{mn}{}^p \rightarrow f'_{mn}{}^p = R_m{}^q R_n{}^r 
  (R^{-1})_s{}^p f_{qr}{}^s \,: \qquad
  \begin{cases}
    Q^{mn} \rightarrow \det(R) ((R^{-1})^T Q R^{-1}))^{mn} 
\,, \\
    a_m \rightarrow R_m{}^n a_n \,.
  \end{cases}
\label{structtransf}
\end{align}
These transformations are naturally divided into two complementary sets. First 
there is the group of automorphism transformations with $f_{mn}{}^p = 
f'_{mn}{}^p$, whose dimension is given in table \ref{3Dalgebras} for the 
different algebras and which are described in \cite{refh}. Then there are the 
transformations that change the structure constants, and these can always be 
used \cite{Wald:1984rg, Schirmer:1995dy} to transform $Q^{pq}$ into a diagonal 
form and $a_q$ to have only one component. We will explicitly go through the 
argument.

Consider an arbitrary symmetric matrix $Q^{mn}$ with eigenvalues $\lambda_m$ and 
orthogonal eigenvectors $\vec{u}_m$. Taking 
\begin{align}
  R^T = (\sqrt{d_2 d_3} \, \vec{u}_1, \sqrt{d_1 d_3} \, \vec{u}_2, \sqrt{d_1 d_2} \, \vec{u}_3) \,, 
\label{diagonalisation}
\end{align}
with $d_m \neq 0$ and ${\rm sgn}(d_1) = {\rm sgn}(d_2) = {\rm sgn}(d_3)$
we find that 
\begin{equation}
 Q^{mn} \rightarrow {\rm diag}(d_1 \lambda_1, d_2 \lambda_2, d_3 \lambda_3)\, .
\end{equation}
We now distinguish between four cases, depending on the rank of $Q^{mn}$:
\begin{itemize}
\item Rank$(Q^{mn})=3$: in this case all components of $a_m$ necessarily 
vanish (due to the Jacobi 
identity), and we can take $d_m = \pm 1 / |\lambda_m|$ to obtain 
\begin{align}
  Q^{mn} = \pm {\rm diag}({\rm sgn}(\lambda_1), {\rm sgn}(\lambda_2), 
{\rm sgn}(\lambda_3)) \,, 
  \qquad a_m=(0,0,0) \,.
\end{align}
\item Rank$(Q^{mn})=2$: in this case one eigenvalue vanishes which we 
take to be $\lambda_1$. 
Then we set  $d_i = \pm 1 / |\lambda_i|$, with $i=2,3$, to obtain 
$Q^{mn} = \pm {\rm diag}(0, {\rm sgn}(\lambda_2), {\rm sgn}(\lambda_3))$. 
From the Jacobi identity, it then follows that $a_m = (a,0,0)$.
We distinguish between vanishing and non-vanishing vector.
In the case $a \neq 0$, one might think that one can use $d_1$ to set 
$a=1$, but from the transformation rule of $a_m$ \eqref{structtransf}
and the form of $R$ \eqref{diagonalisation} it can be seen 
that $a \sim \sqrt{d_2 d_3}$, and therefore $a$ can not be fixed by $d_1$. 
In this case we thus have a one-parameter family of Lie algebras:
\begin{align}
  Q^{mn} = \pm {\rm diag}(0, {\rm sgn}(\lambda_2), {\rm sgn}(\lambda_3)) 
\,, \qquad 
  \begin{cases}
    a_m = (0,0,0) \,, \\
    a_m = (a,0,0) \,.
  \end{cases}
\label{oneparameter}
\end{align}
\item Rank$(Q^{mn})=1$: in this case two eigenvalues vanish, 
e.g.~$\lambda_1 = \lambda_2 = 0$. 
We set $d_3 = \pm 1 / |\lambda_3|$ to obtain $Q^{mn} = \pm 
{\rm diag}(0, 0, {\rm sgn}(\lambda_3))$.
Again one distinguishes between $a_m = 0$ and $a_m \neq 0$. 
In the latter case one is left with a vector $a_m = (a_1,a_2,0)$, of which 
$a_1 \sim \sqrt{d_2 d_3}$ and 
$a_2 \sim \sqrt{d_1 d_3}$.
Thus, one can use $d_1$ and $d_2$ 
to adjust the length of $\vec{a}$ to $1$, after which an 
$O(3)$ transformation in the $(1,2)$-subspace gives 
the final result:
\begin{align}
  Q^{mn} = \pm {\rm diag}(0, 0, {\rm sgn}(\lambda_3)) \,, \qquad 
  \begin{cases}
    a_m = (0,0,0) \,, \\
    a_m = (1,0,0) \,.
  \end{cases}
\end{align}
\item Rank$(Q^{mn})=0$: in this case all three eigenvalues vanish and therefore  
$Q^{mn} = 0$.
Thus, the transformation with matrix \eqref{diagonalisation} is irrelevant.
For $a_m \neq 0$, it follows from \eqref{structtransf} that one can first do 
a scaling to get $|\vec{a}|=1$ and then an $O(3)$ transformation to obtain:
\begin{align}
  Q^{mn} = {\rm diag}(0,0,0) \,, \qquad 
  \begin{cases}
    a_m = (0,0,0) \,, \\
    a_m = (1,0,0) \,.
  \end{cases}
\end{align}
\end{itemize}
Thus, we find that the most general three-dimensional Lie algebra can be 
described by 
$Q^{mn} = \tfrac{1}{2} \text{diag}(q_1,q_2,q_3)$ and $a_m=(a,0,0)$. 
In this basis the commutation relations take the form
\begin{equation}
   [T_1 , T_2] = \tfrac{1}{2} q_3 T_3 -a T_2 \,, \qquad
   [T_2 , T_3] = \tfrac{1}{2} q_1 T_1 \,, \qquad
   [T_3 , T_1] = \tfrac{1}{2} q_2 T_2 +a T_3 \,.
\label{commutations}
\end{equation}
The different three-dimensional Lie algebras are obtained by taking different 
signatures of $Q^{mn}$ and are given in table \ref{3Dalgebras}. Na\"\i vely 
one might conclude that the classification as given above leads to ten 
different algebras. However, it turns out that one has to treat the subcase 
$a=1/2$ of \eqref{oneparameter} as a separate case. We will come back to this 
case below when we discuss the isometries of the group manifold. Thus, the 
total number of inequivalent three-dimensional Lie algebras is eleven, two of 
which are one-parameter families.

\begin{table}[ht]
\begin{center}
\begin{tabular}{||c||c|c||c|c|c|c||}
\hline \rule[-3mm]{0mm}{8mm}
  Bianchi & $a$ & $(q_1,q_2,q_3)$ & Class & Algebra & Dim(Aut) & Dim(Iso) \\
\hline \hline \rule[-3mm]{0mm}{8mm}
  I & 0 & $(0,0,0)$ & A & $u(1)^3$ & $9$ & $6$ \\
\hline \rule[-3mm]{0mm}{8mm}
  II & 0 & $(0,0,1)$ & A & $heis_3$ & $6$ & $4$ \\
\hline \rule[-3mm]{0mm}{8mm}
  III & $\tfrac{1}{2}$ & $(0,-1,1)$ & B & & $4$ & $4$ \\
\hline \rule[-3mm]{0mm}{8mm}
  IV & 1 & $(0,0,1)$ & B & & $4$ & $3$ \\
\hline \rule[-3mm]{0mm}{8mm}
  V & 1 & $(0,0,0)$ & B & & $6$ & $6$ \\
\hline \rule[-3mm]{0mm}{8mm}
  VI$_0$ & 0 & $(0,-1,1)$ & A & $iso(1,1)$ & $4$ & $3$ \\
\hline \rule[-3mm]{0mm}{8mm}
  VI$_a$ & $a$ & $(0,-1,1)$ & B & & $4$ & $3$ \\
\hline \rule[-3mm]{0mm}{8mm}
  VII$_0$ & 0 & $(0,1,1)$ & A & $iso(2)$ & $4$ & $3, 6$ \\
\hline \rule[-3mm]{0mm}{8mm}
  VII$_a$ & $a$ & $(0,1,1)$ & B & & $4$ & $3,6$ \\
\hline \rule[-3mm]{0mm}{8mm}
  VIII & 0 & $(1,-1,1)$ & A & $so(2,1)$ & $3$ & $3,4$ \\
\hline \rule[-3mm]{0mm}{8mm}
  IX & 0 & $(1,1,1)$ & A & $so(3)$ & $3$ & $3,4, 6$ \\
\hline
\end{tabular}
\caption{\label{3Dalgebras}\it The Bianchi classification of three-dimensional 
Lie algebras in terms of the components of their structure constants. Note 
that there are two one-parameter families VI$_a$ and VII$_a$ with special 
cases VI$_0$, VII$_0$ and VI$_{a=1/2}$=III. The algebra $heis_3$ denotes the 
three-dimensional Heisenberg algebra. The table also gives the dimensions of 
the automorphism groups and the dimensions of the possible isometry groups of 
the corresponding group manifolds. The identifications in column 5 can be 
found in \cite{refj}.}
\end{center}
\end{table}

Of the eleven Lie algebras, 
 only $SO(3)$ and $SO(2,1)$ are simple while the rest are all 
non-semi-simple \cite{Schirmer:1995dy,Hamermesh}.
In the non-semi-simple cases we always have $q_1=0$, for which choice the 
Abelian invariant subgroup is $\{ T_2,T_3 \}$, since $T_1$ does not appear on 
the right-hand side in \eqref{commutations}.
All algebras of class A with traceless structure constants fall in the 
$CSO(p,q,r)$-classification with $p+q+r=3$ as discussed in 
\cite{AlonsoAlberca:2003jq} and can give rise to compact and non-compact groups,
while all algebras of class B correspond to non-compact groups \cite{Schirmer:1995dy}.

In addition to the different three-dimensional Lie groups, one can consider 
their realisations as (a subgroup of) the isometry groups of 
three-dimensional Euclidean manifolds. It is well established \cite{Bianchi} 
that, given an $n$-dimensional simply transitive group (which all the groups 
corresponding to type I up to IX are), there is a corresponding 
$n$-dimensional manifold that allows this group as isometries. This manifold 
is called the group manifold. The manifold has by definition at least $n$ 
isometries whose right-invariant Killing vectors $X_a = X_a{}^b \partial / 
\partial z_b$ with $a,b=1,\ldots,n$ satisfy
\begin{align}
  [ X_a , X_b ] = - f_{ab}{}^c X_c \,.
\label{rKilling}
\end{align}
The full group of isometries may very well be bigger.

Let us first consider the case $n=2$, i.e.~two-dimensional manifolds.
The isometry groups of surfaces are zero-, one- or three-dimensional 
\cite{Bianchi}.
Thus, if one requires a two-dimensional simply transitive group to be 
realised as isometries on a two-dimensional manifold, one finds isometry 
enhancement: the full isometry group is necessarily three-dimensional.
Thus every two-dimensional group manifold has the maximum number of isometries
and therefore constant curvature.

Turning to the case $n=3$, the dimension of the isometry group, Dim(Iso), is 
restricted to $0,1,2,3,4$ or $6$ \cite{Bianchi,refa,refc,Szafron}. Since we 
only consider group manifolds, on which a three-dimensional simply transitive 
group is realised as isometries, we have Dim(Iso) $\geq 3$. The three 
right-invariant Killing vectors \eqref{rKilling} corresponding to these 
isometries are given in \eqref{isometries} and \eqref{isometries2} (for our 
parameterization of the structure constants \eqref{commutations}). However, 
the full isometry group of the manifold may well be bigger. Consider as an 
example of such an isometry enhancement the group of Bianchi type I, i.e.~the 
translation group in three dimensions. Its generators are the translational 
isometries of a flat manifold. The full isometry group of such a manifold is 
the six-dimensional $ISO(3)$ group. Thus the isometry group of the Bianchi 
type I group manifold is always six-dimensional.

In table \ref{3Dalgebras} we give the dimension of the possible full isometry 
group of the group manifolds for all Bianchi types. For the simple Lie groups, 
i.e.~the ones of type VIII and type IX, and for the non-semi-simple Lie groups 
of type ${\rm VII}_0$ and type ${\rm VII}_a$, there are different 
possibilities depending on the choice of the three-dimensional manifold, 
i.e.~one can have isometry enhancement. Note that the group manifolds of type 
I, V, ${\rm VII}_0$, ${\rm VII}_a$ and IX allow for the maximum number of six 
isometries, in which case one is dealing with a manifold of constant 
curvature. For the one-parameter family of Lie algebras of type VI${}_a$ one 
has isometry enhancement for the value $a=1/2$, which is the reason why it is 
treated as a separate case, i.e.~type III = type VI${}_{a=1/2}$. We will come 
back to the number of isometries of the group manifolds when discussing 
explicit solutions in section \ref{solutions}.

\section{Reduction over a 3D Group Manifold}
\label{SSreduction}

In this section we review the reduction of $D=11$ supergravity over a 
three-dimensional group manifold, leading to gauged supergravities in eight 
dimensions. We will follow \cite{AlonsoAlberca:2003jq} with
 emphasis on the new features when dealing with the Bianchi class B groups. 
To be precise, we get corrections proportional to the parameter $a$ 
(the trace of the structure constants) to the supersymmetry transformation rules, 
which will be important when searching for solutions. 
This is discussed in more detail in section \ref{solutions}. 

The reduction Ansatz is formally the same for class A and B and it involves the following fields
\begin{align}\nonumber
\rm{11D:~~~}&
\left\{\hat{e}_{\hat{\mu}}{}^{\hat{a}},
\hat{C}_{\hat{\mu}\hat{\nu}\hat{\rho}},
\hat{\psi}_{\hat{\mu}}
\right\}\, ,\\
{\rm{8D:~~~}}& 
 \{ e_\mu{}^a, L_m{}^i, \varphi, 
  \ell, A^m{}_\mu, V_{\mu\, mn}, B_{\mu\nu\, m}, C_{\mu\nu\rho}, \psi_\mu, \lambda_i \} \,,
\end{align}
where the indices are defined according to an $8+3$ split of the
11-dimensional space-time: $x^{\hat{\mu}} = (x^\mu, z^m)$ with
$\mu=(0,1,\ldots,7)$ and $m=(1,2,3)$. Space-time indices are written like 
$\hat{\mu} = (\mu, m)$ while the tangent
indices are $\hat{a} = (a, i)$. The three-dimensional space is taken to be a group
manifold and we reduce over its three (non-Abelian) isometries.

Using a particular Lorentz frame, 
the reduction Ansatz for the 11-dimensional bosonic fields is
\begin{equation}\label{Vielbein}
  \hat{e}_{\hat{\mu}}{}^{\hat{a}}   = 
\left(
\begin{array}{cr}
e^{-\frac{1}{6}\varphi} e_{\mu}{}^{a} & 
e^{\frac{1}{3}\varphi} L_{m}{}^{i}A^{m}{}_{\mu} \\
&\\
0             & 
e^{\frac{1}{3}\varphi}L_{n}{}^{i}\,U^{n}{}_{m}   \\
\end{array}
\right) \, 
\end{equation}
and 
\begin{equation}\label{ansatzC}
\hat{C}_{abc} =  e^{\frac{1}{2}\varphi}\, C_{abc}\, ,
\hspace{.3cm}
\hat{C}_{abi} =  L_{i}{}^{m}B_{ab\,m}\, ,
\hspace{.3cm}
\hat{C}_{aij} = 
 e^{-\frac{1}{2}\varphi}\, L_{i}{}^{m} L_{j}{}^{n}\, 
V_{a\, mn} \, ,
\hspace{.3cm} \hat{C}_{ijk}= e^{-\varphi}\epsilon_{ijk} \ell \, .
\end{equation}
The reduction Ansatz
 for the fermions, including the supersymmetry parameter $\hat\epsilon$,
reads as follows:
\begin{equation}
 \begin{array}{rcl}
{\hat \psi}_{\hat a} & = e^{\varphi/12}\left( \psi_{a} - \frac{1}{6} 
\Gamma_{a} \Gamma^{i} \lambda_i \right ) \,, \qquad {\hat\psi}_{i} = 
e^{\varphi/12} \lambda_i \,, \qquad \hat \epsilon = e^{-\varphi/12} \epsilon 
\, .
  \end{array}
\end{equation}
The matrix $L_m{}^i$ describes the five-dimensional $SL(3,\mathbb{R}) / SO(3)$
scalar coset space. 
It transforms under a global $SL(3,\mathbb{R})$ acting from 
the left and a local $SO(3)$ symmetry acting from the right.  We take the
following explicit representative, thus gauge fixing the local 
$SO(3)$ symmetry:
\begin{align}
  L_m{}^i = \left(
  \begin{array}{ccc}
    e^{-\sigma/\sqrt{3}} & 
    e^{-\phi/2+\sigma/2\sqrt{3}} \chi_1 & e^{\phi/2+\sigma/2\sqrt{3}} \chi_2 \\
    0 & e^{-\phi/2+\sigma/2\sqrt{3}} & e^{\phi/2+\sigma/2\sqrt{3}} \chi_3 \\
    0 & 0 & e^{\phi/2+\sigma/2\sqrt{3}}
\end{array}
\right) \,,
\label{Lscalar}
\end{align}
which contains two dilatons, $\phi$ and $\sigma$, and three 
axions $\chi_m$. It is convenient to define the local $SO(3)$ invariant scalar matrix
\begin{equation}
  \mathcal{M}_{mn} = - L_m{}^i L_n{}^j \eta_{ij} \, ,
\label{Mscalar}
\end{equation} 
where $\eta_{ij} = -\mathbb{I}_3$ is the internal flat metric.

The only internal coordinate dependence in the Ansatz appears via the matrix 
$U^m{}_n$, which is defined in terms of the left-invariant Maurer-Cartan 
1-forms of a 3-dimensional Lie group
\begin{equation}
\sigma^{m}\equiv U^{m}{}_{n}dz^{n}\,.
\label{MaurerCartan}
\end{equation}
By definition these 1-forms satisfy the Maurer-Cartan equations
\begin{equation}
  d\sigma^{m} =-{\textstyle\frac{1}{2}}
  f_{np}{}^{m}\sigma^{n}\wedge \sigma^{p} \,, \qquad
  f_{mn}{}^{p} = -2(U^{-1})^{r}{}_{m} (U^{-1})^{s}{}_{n}\, 
 \partial_{[r} U^{p}{}_{s]}\, ,
\label{MC}
\end{equation}
where the $f_{mn}{}^{p}$ are independent of $z^m$ and form the structure 
constants of the group manifold. Note that we use a slight extension of the 
original procedure of Scherk and Schwarz \cite{Scherk:1979zr} by allowing for 
structure constants with non-vanishing trace (leading to class B
supergravities). This 
corresponds to a group manifold which does not have a constant volume-element. 
We find, by explicitly performing the group manifold procedure, that
the class B reduction can be performed on-shell, i.e.~at the level of the equations
of motion or the supersymmetry variations, but not at the level of the action.
Indeed, the lower-dimensional field equations can not be integrated to an action.

An explicit representation of the Maurer-Cartan 1-forms for general rank
 of the matrix $Q$ was given in \cite{AlonsoAlberca:2003jq} in the case of 
class A. Including class B, i.e.~$a \ne 0$, leads to the following matrix
\begin{align}
  U^m{}_n = \left(
  \begin{array}{ccc}
    1 & 0 &-s_{1,3,2} \\
    0 & e^{a z^1}\,c_{2,3,1} & e^{a z^1}\,c_{1,3,2} \, s_{2,3,1} \\
    0 & -e^{a z^1}\,s_{3,2,1} & e^{a z^1}\,c_{1,3,2} \, c_{2,3,1}
  \end{array} \right) \,,
\label{explicitU}
\end{align}
where we have used the following abbreviations 
\begin{align}
  c_{m,n,p} \equiv \cos(\tfrac{1}{2} \sqrt{q_m}\sqrt{q_n} \, z^p) \,, \qquad
  s_{m,n,p} \equiv \sqrt{q_m} \sin(\tfrac{1}{2}\sqrt{q_m}\sqrt{q_n} \, z^p) /\sqrt{q_n} \, ,
\end{align}
and it is understood that the structure constants satisfy the Jacobi identity, amounting to 
$q_1 a =0$. 

The relation between the Maurer-Cartan 1-forms $\sigma^m$ and the 
three-dimensional isometry groups is as follows.
The metric on the group manifold reads
\begin{align}
  ds^2 = e^{2 \varphi /3} \mathcal{M}_{mn} \sigma^m \sigma^n \,,
\label{bs}
\end{align}
where the scalars $\varphi$ and $\mathcal{M}$ are constants from the three-dimensional point of view. 
A vector field $X$ defines an isometry if it leaves the metric invariant
\begin{align}
{\cal L}_{{\scriptscriptstyle X}}g_{mn}=0\,.
\end{align}
For all values of the scalars, the group manifold has three isometries generated by the right invariant Killing vector fields.
These fulfil the stronger requirement
\begin{align}
  {\cal L}_{{\scriptscriptstyle X}_m}\sigma^n= 0 \label{lsigma}
\end{align}
for all three Maurer-Cartan forms on the group manifold and generate the algebra as given in \eqref{rKilling}.
In the class A case, i.e.~$a=0$, the right-invariant Killing vectors generating the three isometries
are given by
\begin{align}\label{isometries}
X_1 &=\frac{c_{1,2,3}}{c_{1,3,2}}\,\frac{\partial}{\partial z^1}-s_{2,1,3}\,\frac{\partial}{\partial z^2}+\frac{c_{1,2,3}\,s_{3,1,2}}{c_{1,3,2}}\,\frac{\partial}{\partial z^3}\,,\nonumber\\
X_2 &=\frac{s_{1,2,3}}{c_{1,3,2}}\,\frac{\partial}{\partial 
z^1}+c_{1,2,3}\,\frac{\partial}{\partial 
z^2}-\frac{s_{1,2,3}\,s_{1,3,2}}{c_{1,3,2}}\,\frac{\partial}{\partial 
z^3}\,,\\\nonumber X_3 &=\frac{\partial}{\partial z^3}\,,
\end{align}
whereas in the class B case, i.e.~$q_1 = 0$ and $a \ne 0$, they are given by
\begin{align}\label{isometries2}
X_1 &=\frac{\partial}{\partial z^1}-(a z^2+\tfrac{1}{2}\,q_2 z^3)
\,\frac{\partial}{\partial z^2}+(\tfrac{1}{2}\,q_3 z^2-a z^3)
\,\frac{\partial}{\partial z^3}\,,\nonumber\\
X_2 &=\frac{\partial}{\partial z^2}\,,\hskip 2truecm 
X_3 =\frac{\partial}{\partial z^3}\,.
\end{align}
Here, $\partial / \partial z^2$ and $\partial / \partial z^3$ are manifest 
isometries. This follows from the fact that the matrix $U^n{}_m$ is 
independent of $z^2$ and $z^3$.

With the Ansatz above, class B gauged supergravities can be obtained. For our 
present purposes, it is enough to reduce the supersymmetry transformation 
rules. Since we are primarily interested in domain wall solutions, we will 
truncate the reduction to just include the following fields: $g_{\mu\nu}$, 
$L_m{}^i$ and $\varphi$. The resulting $D=8$ fermionic transformations are
\begin{align}\label{ss}
\delta \psi_\mu & = 2 \partial_\mu \epsilon 
-\tfrac{1}{2} \slashed{\omega}_\mu \epsilon  
   +\tfrac{1}{2} \slashed{Q}_\mu \epsilon 
   + \tfrac{1}{24} e^{- \varphi /2} f_{ijk} \Gamma^{ijk} 
\Gamma_\mu \epsilon-\tfrac{1}{6}\,e^{-\varphi/2} f_{ij}{}^{j}\Gamma_\mu 
\Gamma^i\epsilon \, ,\notag \\
\delta \lambda_i & = - \slashed P_{ij}\Gamma^{j}\epsilon 
  - \tfrac{1}{3} \slashed{\partial} \varphi \Gamma_i \epsilon 
  - \tfrac{1}{4} e^{-\varphi/2} (2 f_{ijk} 
-f_{jki}) \Gamma^{jk} \epsilon\, .
\end{align}
Note that there is only one term with an explicit dependence on the trace 
of the structure constants, namely the last term in $\delta \psi_\mu$.
The full supersymmetry rules, without truncation, can be found in appendix 
\ref{ss-rules}.

The global duality group $GL(3,\mathbb{R})$ acts on the indices $m,n,p$ in 
the obvious way and
its action is explicitly given in \cite{AlonsoAlberca:2003jq}. In the gauged 
theory this is in general 
no longer a symmetry since it does not preserve the structure constants. 
The unbroken part is
exactly given by the automorphism group of the structure constants as given 
in table \ref{3Dalgebras}.
Of course it always includes the gauge group, which is embedded in 
$GL(3,\mathbb{R})$ via
\begin{equation}
  g_n{}^m = e^{\lambda^k f_{kn}{}^m}\, ,
\end{equation}
where $\lambda^k$ are the local parameters of the gauge transformations.
However, the full automorphism group can be bigger; for instance it is 
nine-dimensional in the $U(1)^3$ case. Of course this amounts to the fact that 
the ungauged $D=8$ theory has a $GL(3,\mathbb{R})$ symmetry. Note that all 
other cases have Dim(Aut) $< 9$ and thus break the $GL(3,\mathbb{R})$ symmetry 
to some extent. For instance, the $SO(1,1)$ subgroup corresponding to the determinant of 
the $GL(3,\mathbb{R})$ element is broken by all non-vanishing structure constants.

The $GL(3,\mathbb{R})$ transformations are not the only symmetries of the ungauged theory.
There are two more generators leading to the full U-duality group
\begin{align}
  SL(3,\mathbb{R}) \times SL(2,\mathbb{R}) \,. 
\end{align}
The $SL(3,\mathbb{R})$ and the $SO(1,1)$ subgroup of $SL(2,\mathbb{R})$ conspire 
to form the $GL(3,\mathbb{R})$. Its fate after a non-trivial 
gauging has been discussed above, giving rise to the automorphism
groups. To understand the fate of the other subgroups of $SL(2,\mathbb{R})$, 
one needs to define the doublet
\begin{align}
  \vec{f}_{mn}{}^p = \left( \begin{array}{c} f_{mn}{}^p \\ 0 \end{array} \right) \,.
\label{doublet}
\end{align}
Under a global $SL(2,\mathbb{R})$ transformation the full theory is invariant up to a transformation of the structure constants: 
\begin{align}
  \vec{f}_{mn}{}^p \rightarrow \Omega \vec{f}_{mn}{}^p \,, \qquad \Omega \in SL(2,\mathbb{R}) \,.
\end{align}
From this transformation, one can see that the $SO(2)$ and $SO(1,1)$ subgroups of $SL(2,\mathbb{R})$ are broken by any non-zero structure constants\footnote{
  The type II, VI$_0$ and VII$_0$ theories are related via an $SO(2)$ 
  transformation of 90 degrees to the Kaluza-Klein reduction of the $D=9$ gauged theories of 
  \cite{Meessen:1998qm, Bergshoeff:2002mb}. Moreover, after any $SO(2)$ 
  transformation the type II theory can only be further uplifted to the $D=10$ massive
  IIA theory, see e.g.~\cite{Singh:2001gt}.} 
and thus in all theories except the Bianchi type I.
In contrast, the doublet of structure constants \eqref{doublet} is invariant under an $\mathbb{R}$ subgroup of the $SL(2,\mathbb{R})$ symmetry.

\section{Nine-dimensional Origin}
\label{D=9}

In this section we will show that 
all $D=8$ gauged supergravities except those whose gauge group is simple,
i.e.~$SO(3)$ or $SO(2,1)$, can be obtained by a generalised 
reduction of maximal $D=9$ ungauged supergravity 
using its global symmetry group\footnote{This is a different reduction 
Ansatz than 
the group manifold procedure as discussed in section \ref{SSreduction}. 
It is based on
internal rather than space-time symmetries, see also 
\cite{AlonsoAlberca:2003jq} for a discussion.}
 \cite{Scherk:1979ta}. This is possible 
since all these theories follow from the reduction over a non-semi-simple
group manifold with two commuting isometries.
If these two isometries are manifest, as in \eqref{explicitU} with 
$q_1=0$, one can first perform a 
Kaluza-Klein reduction over $T^2$ to nine dimensions.

Restricting ourselves to only those symmetries
that are not broken by $\alpha^\prime$-corrections, the $D=9$
global symmetry group is given by
\begin{equation}\label{dualitygroup}
SL(2,\mathbb{R}) \times SO(1,1)\,.
\end{equation}
Here the duality group  $SL(2,\mathbb{R})$ is a symmetry of the action
and is not violated by $\alpha^\prime$-corrections, since
it descends from the duality group $SL(2,\mathbb{R})$ of
type IIB string theory. We denote its elements by $\Omega$.
The explicit $SO(1,1)$ with elements $\Lambda$ is a symmetry of the equations 
of motion only. Since it has an M-theory origin
as the scaling symmetry\footnote{
In the notation of \cite{Bergshoeff:2002nv}, this corresponds to the 
combination $\alpha - \tfrac{3}{4} \delta$.} 
$x^{\hat \mu} \rightarrow \Lambda \, x^{\hat \mu}$ for ${\hat \mu}=10,11$, this is not violated
by $\alpha^\prime$-corrections either. This $SO(1,1)$ is precisely
the scale transformation with parameter $\Lambda = \exp(a z^1)$,
generated by the matrix $U^m{}_n$, see \eqref{explicitU}, for $q_1=q_2=q_3=0$.
Note that this scaling symmetry scales the volume-element of the torus, which explains
why it is only a symmetry of the $D=9$ equations of motion.

We now perform a $D=9$ to $D=8$ Scherk-Schwarz reduction with fluxes 
\cite{Scherk:1979ta}, making use of (combinations of) the global symmetries 
discussed above. We distinguish between the cases where $\Lambda = 1\ (a=0)$ 
and where  
$\Lambda \neq 1\ (a\ne 0)$. Furthermore, we allow $\Omega$ to be 
either the 
identity or an element of the three subgroups of $SL(2,\mathbb{R})$. Reduction 
to $D=8$ thus gives rise to eight different possibilities, one of which has to 
be split in two. These correspond to the nine $D=8$ maximal gauged 
supergravities with non-semi-simple gauge groups, i.e.~all Bianchi types 
except type VIII with gauge group $SO(2,1)$ and type IX with gauge group 
$SO(3)$. The result is given in table \ref{ninedimred}.

\begin{table}[ht]
\begin{center}
\hspace{-1cm}
\begin{tabular}{||c||c|c||}
\hline \rule[-1mm]{0mm}{6mm}
  $D=9 \Rightarrow D=8$ & $\Lambda = 1$ & $\Lambda \neq 1$ \\
  Reduction Ansatz & ($\Rightarrow$ class A) & ($\Rightarrow$ class B) \\
\hline \hline \rule[-1mm]{0mm}{6mm}
  $\Omega = \mathbb{I}_2$ & I $= U(1)^3$ & V \\
\hline \rule[-1mm]{0mm}{6mm}
  $\Omega \in \mathbb{R}$ & II $= {\rm Heis}_3$ & VI \\
\hline \rule[-1mm]{0mm}{6mm}
  $\Omega \in SO(1,1)$ &  VI$_0$ $= ISO(1,1)$ & III = VI$_{a=1/2}$, VI$_a$ \\
\hline \rule[-1mm]{0mm}{6mm}
  $\Omega \in SO(2)$ & VII$_0$ $= ISO(2)$ & VII$_a$ \\
\hline
\end{tabular}
\caption{\it The different non-semi-simple Bianchi types of 
$D=8$ gauged supergravities, resulting from 
reduction of $D=9$ ungauged supergravity by using different combinations 
of subgroups of the global symmetry groups in $D=9$. Here $\Omega$ and 
$\Lambda$ denote elements of $SL(2,\mathbb{R})$ and $SO(1,1)$, respectively.
\label{ninedimred}}
\end{center}
\end{table}

It can be seen that class A gauged supergravities are obtained by using only a 
subgroup of $SL(2,\mathbb{R})$, which is a reduction that can be performed on 
the $D=9$ ungauged action. Class B gauged supergravities, however, require the 
use of the extra $SO(1,1)$ symmetry which indeed can only be performed at the 
level of the field equations. The connection with $D=9$ clearly shows how 
it is possible to obtain the theories of class B from higher dimensions.

\section{Domain Wall Solutions}\label{solutions}

In this section we will focus on various solutions to the class A and B 
supergravities in $D=8$ and also discuss the uplifting of these solutions to 
$D=11$. For the class B supergravities we show that there are no domain wall 
solutions in $D=8$ that preserve any fraction of the supersymmetry. We do, 
however, find a cosmologically interesting
(non-supersymmetric and time-dependent) space-like domain wall solution, 
i.e.~a domain wall solution where the single transverse direction is time.

In \cite{AlonsoAlberca:2003jq}, we obtained the most general half 
supersymmetric domain wall solutions of the class A supergravities:
\begin{align}\label{triple}
  ds^2 & = H^{\frac{1}{12}}dx_7^2-H^{-\frac{5}{12}}dy^2\,, \nonumber \\
  e^\varphi & = H^{\frac{1}{4}}, \hspace{0.5cm} 
  e^\sigma = H^{-\frac{1}{2\sqrt{3}}}h_1^{\frac{\sqrt{3}}{2}}, \hspace{0.5cm}
  e^\phi= H^{-\frac{1}{2}}h_1^{\frac{1}{2}} h_2 \,, \\
  \chi_1 & = \chi_2 = \chi_3 = 0 \,, \notag
\end{align}
where the dependence on the transverse coordinate $y$ is governed by
\begin{align}
& H(y) = h_1 h_2 h_3 \,, \qquad h_1 \equiv q_1y+c_1, \qquad h_2 \equiv q_2y+c_2, \qquad h_3\equiv q_3y+c_3 \,. \label{harmonics}
\end{align}
Here $c_m$ are arbitrary constants whose values will affect the 
range of $y$, due to the obvious requirement $h_m > 0$. The Killing spinor 
satisfies the condition
\begin{align}
  (1+ \Gamma_{y123}) \epsilon = 0 \, ,
\end{align}
where the indices $1,2,3$ refer to the internal group manifold directions.
Note that the dependence on the transverse coordinate $y$ is expressed in 
terms of three functions $h_m$ which are harmonic on $\mathbb{R}$.  
We define $n$ to be the number of linearly independent harmonics
$h_m$ with $q_m \neq 0$.
The maximal value of $n$ in a specific class is then given by the number 
of non-zero $q_m$'s of the corresponding structure constants.
We call the solution an $n$-tuple domain wall\footnote{
  Compare this to e.g.~the D8-brane which is expressed in terms 
  of one harmonic function, $h=1+m y$, where the 
  mass parameter $m$ is piecewise constant. The domain walls are 
  located at the points in $y$ where $m$ is discontinuous. In the same way, 
  our $n$ constituent domain walls will be located where the corresponding
  $q_m$ change values. In \cite{Bergshoeff:2003sy}, the double domain wall of \cite{Bergshoeff:2002mb} 
  is given in a form similar to (\ref{triple}).} with $n \leq 3$.
In this terminology, $n=3$ gives a triple, $n=2$ a 
double and $n=1$ a single domain wall, while
$n=0$ is flat space-time \cite{AlonsoAlberca:2003jq}.

Note that the solution \eqref{triple} is given in an $SL(3,\mathbb{R})$ frame where
the three-dimensional gauge freedom has been fixed. The solution for all gauge choices is 
given in \cite{AlonsoAlberca:2003jq}. 
In addition to the gauge group, one can use the
larger automorphism group (of which the gauge group with constant parameters is a subgroup) 
to set $c_m = 1$ if $q_m = 0$. Furthermore, one parameter
can be set to zero by shifting the transverse coordinate $y$.
Thus, the number of parameters of the solution is $n-1$ (for $n\ge 1$).

Upon uplifting to $D=11$, using the relation (\ref{Vielbein}), we 
find that the $n$-tuple domain wall solutions become purely gravitational 
solutions with a metric of the form $\hat{ds}{}^2 =dx_7{}^2-ds_4{}^2$, where
\begin{align}
  ds_4{}^2 = H^{-\frac{1}{2}}dy^2+H^{\frac{1}{2}}
  \left ( \frac{\sigma_1^2}{h_1}+\frac{\sigma_2^2}{h_2}+\frac{\sigma_3^2}{h_3} 
\right) \,. \label{4dmetric}
\end{align}
Here $\sigma^m$ are the Maurer-Cartan 1-forms defined in \eqref{MaurerCartan} 
and \eqref{explicitU}. The uplifted solutions are all 1/2 BPS except for the 
cases when $h_1=h_2=h_3$ (only possible for Bianchi I and IX), which uplift to 
flat spacetime and thus become fully supersymmetric upon uplifting. Note that 
the solution \eqref{4dmetric} is an extension to different Bianchi types of 
the generalised Eguchi-Hanson solution constructed in \cite{Belinskii:1978}.

We would like to see whether there are also supersymmetric domain wall 
solutions to the class B supergravities. It turns out that for this case there 
are no domain wall solutions preserving any fraction of supersymmetry. This 
can be seen as follows. The structure of the BPS equations requires the 
projector for the Killing spinor of a 1/2 BPS domain wall solution to be the 
same as above. The presence of the extra term in $\delta \psi_\mu$ (see 
\eqref{ss}), depending on the trace of the structure constants, implies that 
there are no domain wall solutions with this type of Killing spinor, since the 
structure of $\Gamma$-matrices of this term cannot be combined with other 
terms. To get a solution, one is forced to put $f_{ij}{}^j=0$, thus leading 
back to the class A case. This also follows from $\delta\lambda_i$, since the 
resulting equation is symmetric in two indices, except for a single 
antisymmetric term, containing $f_{ij}{}^j$. Next, we search for domain wall 
solutions preserving an arbitrary fraction of the supersymmetry. From the 
structure of the BPS equations, it is seen that only one additional kind of 
projector is allowed, namely
\begin{align}
  (1+\Gamma_{\alpha 123})\,\epsilon=0
\end{align}
where $\alpha\neq y$ and space-like. However, this again leads to $f_{ij}{}^j=0$. 
We conclude that there are no domain wall solutions preserving any fraction of supersymmetry for 
the class B supergravities.

As we have shown in section \ref{SSreduction}, the internal three-dimensional 
manifolds are by definition invariant under the three-dimensional group 
of isometries given in \eqref{isometries} and \eqref{isometries2}. 
This holds for arbitrary values of the scalars in \eqref{bs}.
However, there can be more isometries, that rotate two of the Maurer-Cartan 
one-forms $\sigma^m$ and $\sigma^p$ into each other. This is an isometry of the 
metric in two cases:
\begin{itemize}
\item $q_m = q_p = 0$: In this case one can use the automorphism group 
to set $c_m = c_p = 1$.
Equation \eqref{4dmetric} shows that a rotation between $\sigma^m$ and 
$\sigma^p$ is an isometry for all solutions of this class.
\item $q_m = q_p \neq 0$: In this case one must set $c_m = c_p$ by hand, 
after which a 
rotation between $\sigma^m$ and $\sigma^p$ is an isometry. Thus, this only 
holds for a truncation of the solutions of this class and since $h_m=h_p$  
corresponds to decreasing $n$ by one.
\end{itemize}
This leads to the different possibilities summarised in table 
\ref{isometryenhancement}. Note that these exhaust all possible number of 
isometries on three-dimensional class A group manifolds as given in table 
\ref{3Dalgebras}. The extra fourth isometry was constructed by Bianchi 
\cite{Bianchi} for the types II, VIII and IX. He claimed that type VII${}_0$ 
did not allow for isometry enhancement but the existence of three extra 
Killing vectors\footnote{
  We thank Sigbj\o rn Hervik for a valuable discussion on this point.} 
was later shown in \cite{refa,refc,Szafron}. These three extra isometries 
appear upon identifying the two $y$-dependent harmonics. Note that the extra 
isometries may not be isometries of the full manifold in which the group 
submanifold is embedded. Indeed, this is what happens for type VII$_0$ where 
two of the extra isometries are $y$-dependent and therefore do not leave the 
full metric invariant \cite{refa,refc,Szafron}. The extra Killing vectors of 
the group manifold for the uplifted domain wall solutions (\ref{triple}) are 
explicitly given in appendix \ref{killingvectors} for completeness.

\begin{table}[ht]
\begin{center}
\begin{tabular}{||c|c||c|c|c|c||}
\hline \rule[-3mm]{0mm}{8mm}
  Bianchi & $(q_1,q_2,q_3)$ & $n=0$ & $n=1$ & $n=2$ & $n=3$ \\
\hline \hline \rule[-3mm]{0mm}{8mm}
  I & $(0,0,0)$ & $6$ & - & - & - \\
\hline \rule[-3mm]{0mm}{8mm}
  II & $(0,0,1)$ & - & $4$ & - & - \\
\hline \rule[-3mm]{0mm}{8mm}
  VI${}_0$ & $(0,-1,1)$ & - & - & $3$ & - \\
\hline \rule[-3mm]{0mm}{8mm}
  VII${}_0$ & $(0,1,1)$ & - & $6$ & $3$ & - \\
\hline \rule[-3mm]{0mm}{8mm}
  VIII & $(1,-1,1)$ & - & - & $4$ & $3$ \\
\hline \rule[-3mm]{0mm}{8mm}
  IX & $(1,1,1)$ & - & $6$ & $4$ & $3$ \\
\hline
\end{tabular}
\caption{\label{isometryenhancement}\it The numbers of isometries of the
three-dimensional group manifold for the different $n$-tuple domain wall 
solutions. For a given type one finds isometry enhancement by decreasing $n$, 
i.e.~upon identifying two harmonic functions $h_m$.}
\end{center}
\end{table}

As we have mentioned above, two of the class A solutions uplift to flat 
spacetime in $D=11$: the Bianchi type IX solutions with $n=1$ and all Bianchi
type I solutions (having $n=0$). In view of the discussion above, we can now 
understand why this happens. One can check that the only way to embed 
three-dimensional submanifolds of zero (for type I) or constant positive (for 
type IX) curvature in four Euclidean Ricci-flat dimensions is to embed them in 
four-dimensional flat space. Indeed, this is exactly what we find: the two 
solutions both have a maximally symmetric group manifold with six isometries 
and hence constant curvature and uplift to flat $D=11$ space-time. 

The type VII$_0$ group manifold can also have six isometries and zero curvature.
For the domain wall solutions above, this can not be embedded in four-dimensional 
flat space due to the $y$-dependence of two of its isometries. Note, however, 
that there is another type VII$_0$ solution with flat geometry and vanishing 
scalars that coincides with the type I solution \eqref{triple} given above\footnote{
  This solution coincides, after an $SO(2)$ rotation of 90 degrees, 
  with the Kaluza-Klein reduction of the Mink$_9$ solution \cite{Gheerardyn:2001jj,Bergshoeff:2002nv} 
  of the $SO(2)$ gauged supergravity in $D=9$.}.
The corresponding group manifold can be embedded in four-dimensional flat space and
indeed this solution uplifts to 11-dimensional Minkowski just as the type I solution.
However, unlike its type I counterpart, the 8-dimensional type VII$_0$ solution 
with flat geometry and vanishing scalars breaks all supersymmetry.

When we include the class B supergravities, we deduce from table 
\ref{3Dalgebras} that there are two more cases with maximally symmetric group 
manifolds, which have constant negative curvature, namely type V and type VII$_a$.
The group manifold can only be embedded in a four-dimensional Ricci-flat 
manifold if the embedding space is flat and Lorentzian. Thus one can expect 
solutions of Bianchi type V and type VII$_a$ that have 6 isometries and 
uplift to flat spacetime in $D=11$. It is interesting to find out how these 
extra solutions look like. By solving the field equations in $D=11$ using the 
Bianchi type V or type VII$_a$ Ansatz with constant coset scalars $\mathcal{M}_{mn}$, we find 
the following (non-supersymmetric and time-dependent) solution in $D=8$
\begin{align}\label{timedepsol}
  ds^2 & = dt^2-t^{2/3} dx_7^2\,, \qquad 
  e^\varphi = \tfrac{9}{4} t^2\,,
\end{align}
where all scalars except $\varphi$ have been put to zero using the 
automorphism groups of the Bianchi type V and type VII$_a$ algebras. 
One can view this as a space-like domain wall, i.e.~a domain wall 
where the single transverse direction is time. There are no 
(non-supersymmetric) static domain wall solutions for constant coset scalars 
$\mathcal{M}_{mn}$.

\begin{figure}[tb]
  \centerline{\epsfig{file=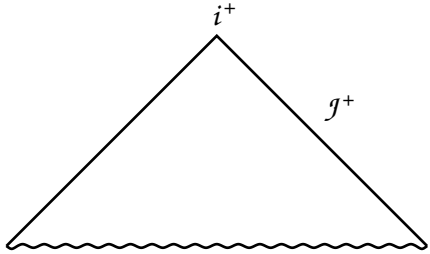,width=.35\textwidth}}
  \caption{\it The Penrose diagram for the solution (\ref{timedepsol}), and for the Einstein-de Sitter universe, is given
   by the upper half of the diamond representing Minkowski spacetime and has a singularity at $t=0$.} \label{PenroseDiag}
\end{figure}

The solution \eqref{timedepsol} describes an expanding universe with the same 
qualitative features as the Einstein-de Sitter universe\footnote{The 
Einstein-de Sitter universe is a flat ($k=0$) matter-dominated ($p=0$) 
Robertson-Walker spacetime with zero cosmological constant ($\Lambda = 0$).}. 
In the present case the stress-energy tensor is generated by the scalar field 
$\varphi$. The $D=8$ Ricci scalar is given by $R = \tfrac{14}{9} \, t^{-2}$. 
Note that the metric \eqref{timedepsol} can be rewritten as being conformal to 
Minkowski spacetime, by the coordinate change $\tau\sim t^{2/3}$. The Penrose 
diagram for the solution \eqref{timedepsol} is therefore given by the upper 
half of the diamond that represents Minkowski spacetime with a singularity at 
$t=0$, see figure \ref{PenroseDiag}.

Upon uplifting to $D=11$ (and rescaling $t$), we get the following solution
\begin{equation}\label{flat}
  d\hat{s}^2= dt^2-dx_7^2- \tfrac{9}{4} t^2 \sigma^m \sigma^m \,,
\end{equation}
where summation over $m$ is understood. The Maurer-Cartan 1-forms $\sigma^m$ 
are defined in \eqref{MaurerCartan} and \eqref{explicitU}. The metric 
\eqref{flat} is a 7D flat Euclidean metric times a 4D metric with Lorentz 
signature which turns out to be a particular parameterization of flat 
spacetime. Thus, we indeed find that the eight-dimensional solution uplifts to 
the maximally (super-)symmetric flat spacetime in $D=11$ as expected. If we 
instead reduce on the seven-dimensional group manifold obtained by taking a 
Bianchi type V or VII$_a$ manifold times 
$T^4$, we get the Einstein-de Sitter universe as a solution\footnote{We have 
to go to a particular frame in order to get exactly the Einstein-de Sitter 
solution in $D=4$. In e.g.~the Einstein frame we instead find a conformally 
related solution. We can in this way also obtain Einstein-de Sitter-like 
solutions for all $D\leq 8$.} in $D=4$. This solution also uplifts to flat 
spacetime in $D=11$. Note that the equations of motion in both $D=8$ and $D=4$ 
can not be obtained from an action. The nice feature of obtaining Einstein-de 
Sitter universes in this way is that they have a very simple and natural 
higher-dimensional origin, namely the only maximally (super-)symmetric vacuum 
solution in $D=11$, i.e.~flat spacetime.

\medskip

\section{Conclusions}\label{conclusions}

In this work we have constructed eleven maximal $D=8$ gauged supergravities in 
terms of the Bianchi classification of three-dimensional Lie groups, which 
distinguishes between class A and B. We find that this distinction carries 
over to a number of features of the eight-dimensional theories. Class A 
theories can be formulated in terms of an action, whereas the theories of 
class B have equations of motion that cannot be integrated to an action. 
Moreover, only the supergravities of class A admit 1/2 BPS domain wall 
solutions. These solutions provide realisations of isometry enhancement in the 
group manifold after identification of the harmonics. The three solutions that 
have a maximum number of isometries uplift to $D=11$ flat spacetime.

We find that there are no domain wall solutions for the class B theories that 
preserve any supersymmetry. However, we have found a (non-supersymmetric and 
time-dependent) space-like domain wall solution to two of the class B 
theories. The solution describes an expanding universe with the same 
qualitative features as the Einstein-de Sitter universe. By instead reducing 
over a seven-dimensional group manifold we obtain the Einstein-de Sitter 
universe as a solution in $D=4$. Both solutions uplift to the only maximally 
(super-)symmetric vacuum solution in $D=11$, i.e.~flat spacetime, which 
provides a nice higher-dimensional origin of Einstein-de Sitter universes.

The Einstein-de Sitter solution has an interesting cosmological 
interpretation. It has recently been argued that Einstein-de Sitter models are 
acceptable models of the universe \cite{Blanchard:2003du}, and e.g.~fit the 
CMB data equally well if not better than the best concordance model. This, 
however, assumes that there must be some other explanation of the observed 
Hubble diagram of distant type Ia supernovae \cite{Riess:1998cb} than a 
positive cosmological constant. It would be interesting to investigate further 
the occurrence of Einstein-de Sitter universes in compactifications of 
M-theory.

\section*{Acknowledgements}

\noindent T.O.~would like to thank the CERN Theory Division for its 
hospitality and financial support and M.M.~Fern\'andez for her continuous 
support. This work is supported in part by the European Community's Human 
Potential Programme under contract HPRN-CT-2000-00131 Quantum Spacetime, in 
which the University of Groningen is associated with the University of 
Utrecht. The work of U.G. is part of the research program of the ``Stichting 
voor Fundamenteel Onderzoek der Materie'' (FOM). 
The work of T.O.~is partially supported by the 
Spanish grant FPA2000-1584.

\appendix
\section{Supersymmetry Rules}
\label{ss-rules}

In this appendix we give the full supersymmetry rules (up to 
higher-order fermions) of all $D=8$ class A and class B
supergravities. 
Considering the Ansatz (\ref{ansatzC}), the dimensional reduction of the 
eleven dimensional field strength $\hat G$ leads to the eight-dimensional field strengths 
\begin{align}\label{curvatures}
G_{\mu \nu \rho \lambda}&= 4\partial_{[\mu}
C_{\nu \rho \lambda ]}+6F^{m}{}_{[\mu \nu}B_{\rho \lambda]\,m}\,,
\nonumber \\
G_{\mu \nu \rho m}&= 
3{\cal D}_{[\mu} B_{\nu \rho ]\,m} + 3 F^{n}{}_{[\mu \nu} V_{\rho]\,mn} \,, \\
G_{\mu \nu mn}&= 2{\cal D}_{[\mu} V_{\nu]\,mn} 
- f_{mn}{}^{p} B_{\mu\nu \,p} + \ell \epsilon_{mnp} F^p{}_{\mu \nu}\,,
\nonumber \\
G_{\mu mnp}&=
\epsilon_{mnp}\partial_\mu \ell +3\left( V_{\mu \,r[ m}+\ell A^q{}_\mu
\epsilon_{qr[m} \right) f_{np]}{}^{r}\, , \nonumber
\end{align}
where the field strength of the gauge field is given by 
\begin{equation}
F^m{}_{\mu \nu }  =  2\partial_{[\mu} A^{m}{}_{\nu]} -
f_{np}{}^{m}A^{n}{}_{\mu}A^{p}{}_\nu \, .
\end{equation}
The curvatures \eqref{curvatures} are invariant under the gauge 
transformations that arise upon reduction of the $D=11$ law
$\delta \hat C_{\hat \mu \hat \nu \hat \rho}=
3\partial_{[\hat \mu}\hat \Lambda_{\hat \nu \hat\rho]}$. 
Using the Ansatz
\begin{align}
\hat \Lambda_{\mu \nu} &= \Lambda_{\mu \nu} - 2 A^m{}_{[\mu} \Lambda_{\nu ]m}
+ A^m{}_{\mu} A^n{}_{\nu} \Lambda_{mn}, \nonumber \\
\hat \Lambda_{\mu m} &= U^q{}_m(\Lambda_{\mu q}-A^n{}_{\mu} \Lambda_{qn}),\\
\hat \Lambda_{mn} &= U^p{}_m U^q{}_n \Lambda_{pq},  \nonumber
\end{align}
the gauge transformations in $D=8$ are
\begin{align}
\delta C_{\mu \nu \rho} &= 3\partial_{[\mu} \Lambda_{\nu \rho]}
-3F^m{}_{[\mu \nu}\Lambda_{\rho ]m} \nonumber , \nonumber \\
\delta B_{\mu \nu\, m}&=2{\cal D}_{[\mu} \Lambda_{\nu]m}
-\Lambda_{mn}F^n{}_{\mu \nu}, \\
\delta V_{\mu\, mn}&={\cal D}_\mu\Lambda_{mn}+f_{mn}{}^{p}
\Lambda_{\mu p} ,\nonumber \\
\delta \ell&=\frac{1}{2}\epsilon^{mnp}f_{mn}{}^{q}\Lambda_{qp} . \nonumber 
\end{align}

The supersymmetry transformation rules in eight dimensions are
\begin{align}
\delta e_{\mu}{}^{a} = &
-\frac{i}{2}\overline{\epsilon} \Gamma^{a} {\psi}_{\mu} \, \notag \\
\delta \psi_\mu  = & 2 \partial_\mu \epsilon 
 - \frac{1}{2} {\slashed \omega}_\mu \epsilon
  +\frac{1}{2}L_{[i|}{}^{m} {\cal D}_{\mu} L_{m|j]}\Gamma^{ij} \epsilon \,
+\frac{1}{24}e^{-\varphi/2}f_{ijk}\Gamma^{ijk} \Gamma_\mu \epsilon 
-\tfrac{1}{6}\,e^{-\varphi/2} f_{ij}{}^{j}\Gamma_\mu \Gamma^i\epsilon
\, \notag \\
& +\frac{1}{24} e^{\varphi/2}\Gamma^iL_{i}^{\ m}
( \Gamma_{\mu}^{\ \nu \rho}-10\delta_\mu^{\ \nu}\Gamma^\rho )
 F_{m\nu \rho}\epsilon 
-\frac{i}{12}e^{-\varphi}\Gamma^{ijk}L_i{}^m L_j{}^n L_k{}^p G_{\mu mnp} 
\epsilon \notag \\
& +\frac{i}{96}e^{\varphi/2}(\Gamma_{\mu}^{\ \nu \rho \delta \epsilon}
-4 \delta^{\ \nu}_{\mu} \Gamma^{\rho \delta \epsilon})
G_{\nu \rho \delta \epsilon} \epsilon 
+ \frac{i}{36}\Gamma^iL_i^{\ m}(\Gamma_{\mu}^{\ \nu \rho \delta}
-6 \delta^{\ \nu}_\mu \Gamma^{\rho \delta})G_{\nu \rho \delta m}\epsilon
\notag \\ 
& +\frac{i}{48}e^{-\varphi/2}\Gamma^i \Gamma^j L_i^{\ m} L_j^{\ n}
(\Gamma_\mu^{\ \nu \rho}-10\delta_\mu^{\ \nu}\Gamma^\rho)
G_{\nu \rho mn} \epsilon \,, \notag \displaybreak[2] \\
\delta \lambda_i  = &
 \frac{1}{2}L_i^{\ m} L^{jn}{\slashed {\cal D}}{\cal M}_{mn}
 \Gamma_j \epsilon -\frac{1}{3} {\slashed \partial} \varphi 
\Gamma_i \epsilon 
-\frac{1}{4}e^{-\varphi/2} (2f_{ijk}-f_{jki})\Gamma^{jk}\epsilon \, \notag \\
& + \frac{1}{4}e^{\varphi/2}L_i^{\ m}{\cal M}_{mn}{\slashed F}^n \epsilon +
\frac{i}{144}e^{\varphi/2}\Gamma_i {\slashed G}\epsilon 
+\frac{i}{36}(2\delta_i^{\ j}-\Gamma_{i}^{\ j})L_j^{\ m}{\slashed G}_m 
\epsilon \notag \\
& +\frac{i}{24}e^{-\varphi/2}\Gamma^j L_j^{\ m}L_k^{\ n}
(3\delta_i^{\ k}-\Gamma_{i}^{\ k}){\slashed G}_{mn} \epsilon 
+\frac{i}{6}e^{-\varphi}\Gamma^{jk} L_i{}^m L_j{}^n L_k{}^p
{\slashed G}_{mnp} \epsilon \,,
\notag \displaybreak[2] \\
\delta A^m{}_{\mu}
= & -\frac{i}{2}e^{-\varphi/2} L_{i}^{\ m} \overline{\epsilon}  (
   \Gamma^{i} \psi_{\mu} -{\Gamma}_{\mu} 
(\eta^{ij}- \frac{1}{6}\Gamma^{i}\Gamma^{j})\lambda_j )  \,, \notag \\
\delta V_{\mu\, mn} = & \varepsilon_{mnp} [-\frac{i}{2}e^{\varphi/2}
L_i^{\ p} \bar \epsilon ( \Gamma^i \psi_\mu
+\Gamma_\mu (\eta^{ij}-\frac{5}{6}\Gamma^i \Gamma^j) \lambda_j )-
\ell\, \delta A^p{}_\mu ]   \,, \notag \displaybreak[2] \\
\delta B_{\mu \nu\, m} = & 
L_m^{\ \ i} \bar \epsilon  (\Gamma_{i[\mu} \psi_{\nu ]}
+\frac{1}{6} \Gamma_{\mu \nu}
(3\delta_i^{\ j}-\Gamma_i \Gamma^j)\lambda_j ) -2\, \delta A^n{}_{[\mu} V_{\nu]\, mn} \,, \notag \displaybreak[2] \\
\delta C_{\mu \nu \rho} = & 
\frac{3}{2}e^{-\varphi/2} \bar \epsilon \Gamma_{[\mu \nu}( \psi_{\rho]}
-\frac{1}{6}\Gamma_{\rho]} \Gamma^i \lambda_i )
-3 \delta A^m{}_{[\mu} B_{ \nu \rho]\,m} \,, \notag \\ 
L^{\ n}_{i}\delta L_n{}_j = & \frac{i}{4}e^{\varphi/2}
  \overline\epsilon  (\Gamma_i \delta_j^{\ k} + \Gamma_j \delta^{\ k}_i-
  \frac{2}{3}\eta_{ij}\Gamma^k )\lambda_k \,,  \notag \\
\delta \varphi 
   = & -\frac{i}{2} \overline{\epsilon} \Gamma^{i} \lambda_i \,, \notag \\
\delta \ell = & -\frac{i}{2}e^{\varphi} \bar \epsilon \Gamma^i \lambda_i \,.
\end{align}

\section{Killing Vectors}
\label{killingvectors}

In this appendix we give the Killing vectors associated with the isometry 
enhancement taking place for some of the domain wall
solutions, as discussed in section 
\ref{solutions}.

\subsection{Class A}
For the class A solutions we denote the extra Killing vectors $X_4$, $X_5$ and 
$X_6$, corresponding to rotations between $\sigma^1$ and $\sigma^2$, $\sigma^1$ 
and $\sigma^3$ and $\sigma^2$ and $\sigma^3$, respectively.
\begin{itemize}
\item
{ Type I} with $Q=\tfrac{1}{2} \, {\rm diag}(0,0,0)$: \ \
\begin{align}
X_4&=-z^2\,\frac{\partial}{\partial z^1}+z^1\,\frac{\partial} {\partial
z^2}\,, \qquad 
X_5=-z^3\,\frac{\partial}{\partial
z^1}+z^1\,\frac{\partial} {\partial z^3}\,, \qquad 
X_6=-z^3\,\frac{\partial} {\partial z^2}+z^2\,\frac{\partial} {\partial 
z^3}\,.
\end{align}
\item
{ Type II} with $Q=\tfrac{1}{2} \, {\rm diag}(0,0,1)$: \ \
\begin{align}
X_4&=-z^2\,\frac{\partial}{\partial z^1}+z^1\,\frac{\partial} {\partial 
z^2}+\tfrac{1}{4}\,((z^1)^2-(z^2)^2)\,\frac{\partial} {\partial z^3}\,.
\end{align}
\item
{ Type VII$_0$} with $Q=\tfrac{1}{2} \, {\rm diag}(0,1,1)$ with 
$h(y)=h_2=h_3$: \ \
\begin{align}
X_4&=-h^{-1/2}z^2\,\frac{\partial} {\partial z^1}+h^{1/2}z^1\,\frac{\partial} 
{\partial z^2}\,, \nonumber\\
X_5&=-h^{-1/2}z^3\,\frac{\partial} {\partial z^1}+h^{1/2}z^1\,\frac{\partial}
{\partial z^3}\,,\\
X_6&=-z^3\,\frac{\partial} {\partial z^2}+z^2\,\frac{\partial} {\partial 
z^3}\,.\nonumber
\end{align}
\item
{ Type VIII} with $Q=\tfrac{1}{2} \, {\rm diag}(1,-1,1)$: \ \ 
\begin{align}
X_5&=\frac{s_{3,2,1}s_{1,3,2}}{c_{1,3,2}}\,\frac{\partial}{\partial 
z^1}+c_{3,2,1}\,\frac{\partial} {\partial 
z^2}+\frac{s_{3,2,1}}{c_{1,3,2}}\,\frac{\partial} {\partial z^3}\,.
\end{align}
\item
{ Type IX} with $Q=\tfrac{1}{2} \, {\rm diag}(1,1,1)$: \ \ 
\begin{align}
X_4&=-\frac{c_{3,2,1}s_{1,3,2}}{c_{1,3,2}}\,\frac{\partial}{\partial 
z^1}+s_{2,3,1}\,\frac{\partial} {\partial
z^2}-\frac{c_{3,2,1}}{c_{1,3,2}}\,\frac{\partial} {\partial z^3}\,,\nonumber\\
X_5&=\frac{s_{3,2,1}s_{1,3,2}}{c_{1,3,2}}\,\frac{\partial}{\partial 
z^1}+c_{3,2,1}\,\frac{\partial} {\partial
z^2}+\frac{s_{3,2,1}}{c_{1,3,2}}\,\frac{\partial} {\partial z^3}\,,\\
X_6&=-\,\frac{\partial}{\partial z^1}\,.\nonumber
\end{align}
\end{itemize}

\subsection{Class B}
For class B there are scalings associated with the non-zero parameter $a$, and 
therefore the extra Killing vectors do not correspond to just rotations among 
the Maurer-Cartan one-forms in this case.
\begin{itemize}
\item
{ Type V} with $Q=\tfrac{1}{2} \, {\rm diag}(0,0,0)$ and $a =1$: \ \
\begin{align}
X_4&=-z^2\,\frac{\partial}{\partial
z^1}+\tfrac{1}{2}((z^2)^2-(z^3)^2-e^{-2 
z^1})\,\frac{\partial}{\partial
z^2}+z^2 z^3\,\frac{\partial}{\partial z^3}\,,\nonumber\\
X_5&=-z^3\,\frac{\partial}{\partial z^1}+z^2 z^3\,\frac{\partial} {\partial 
z^2}+\tfrac{1}{2} ((z^3)^2-(z^2)^2-e^{-2
z^1})\,\frac{\partial} {\partial z^3}\,,\\
X_6&=-z^3\,\frac{\partial} {\partial z^2}+z^2\,\frac{\partial} {\partial 
z^3}\,.\nonumber
\end{align}
\end{itemize}

\bibliography{eight}
\bibliographystyle{utphysmodb}

\end{document}